\documentclass{iaus}
\usepackage{graphicx}

\title[Tidal Evolution of Exoplanets] 
{Tidal Evolution of Close-In Extra-Solar Planets}

\author[Jackson, Greenberg \& Barnes]   
{Brian Jackson$^1$, Richard Greenberg$^1$, \and Rory Barnes$^1$}

\affiliation{$^1$Lunar and Planetary Laboratory, University of Arizona,\\
1629 E University Blvd \\ Tucson, Arizona 85721-0092 USA \\ email: {\tt bjackson@lpl.arizona.edu}}

\pubyear{2008}
\volume{249}  
\pagerange{119--126}
 \jname{Exoplanets: Detection, Formation and
Dynamics} \editors{Y.-S. Sun, S. Ferraz-Mello \& J.-L. Zhou, eds.}
\begin{document}

\maketitle

\begin{abstract}
The distribution of eccentricities $e$ of extra-solar planets with semi-major 
axes $a >$ 0.2 AU is very uniform, and values for $e$ are generally large. 
For $a <$ 0.2 
AU, eccentricities are much smaller (most $e <$ 0.2), a characteristic widely 
attributed to damping by tides after the planets formed and the protoplanetary 
gas disk dissipated. We have integrated the classical coupled tidal evolution 
equations for $e$ and $a$ backward in time over the estimated age of each planet, and confirmed that 
the distribution of initial $e$ values of close-in planets matches that of the 
general population for reasonable tidal dissipation values $Q$, with the best 
fits for stellar and 
planetary $Q$ being $\sim10^{5.5}$ and $\sim10^{6.5}$, respectively. The 
current small values of 
$a$ were only reached gradually due to tides over the lifetimes of the planets, 
\textit{i.e.}, the earlier gas disk migration did not bring all planets to 
their current orbits.  As the orbits tidally evolved, there was substantial tidal heating 
within the planets. The past tidal heating of each planet may have 
contributed significantly to the thermal budget that governed the planet's 
physical properties, including its radius, which in many cases may be measured 
by observing transit events. Here we also compute the plausible heating 
histories for a few planets with anomalously large measured radii, including HD 
209458 b. We show that they may have undergone substantial tidal heating during 
the past billion years, perhaps enough to explain their large radii. Theoretical
models of exoplanet interiors and the corresponding radii should include the role of large and time-variable tidal heating. Our results may have important implications for planet 
formation models, physical models of ``hot Jupiters'', and the success of 
transit surveys.

\end{abstract}


\section{Introduction}
\begin{figure}[b]
\begin{center}
 \includegraphics[width=13cm]{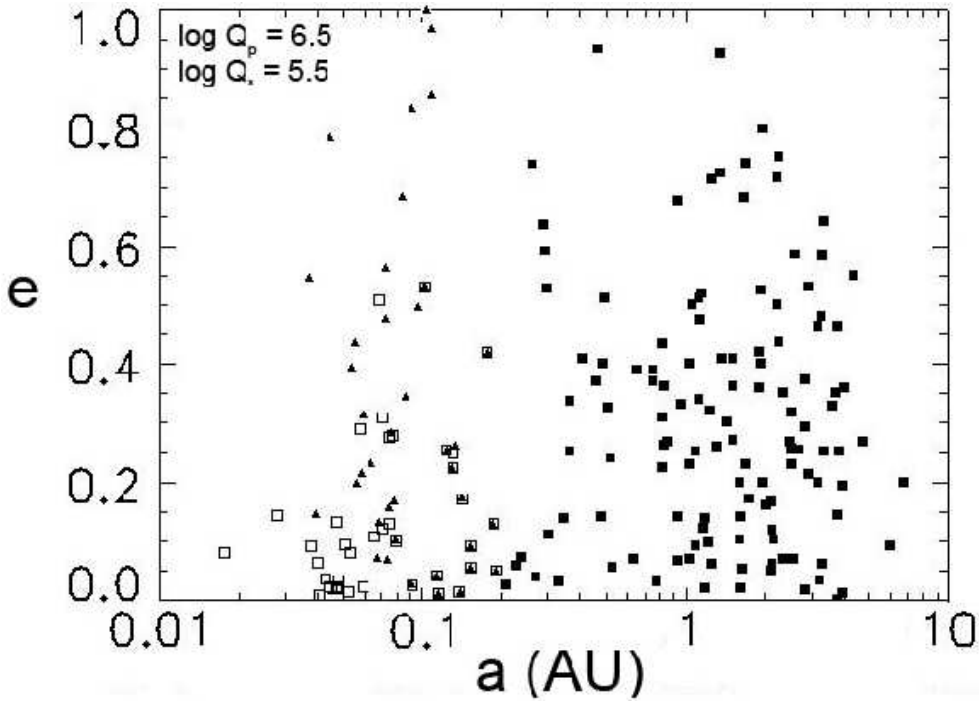}
 \caption{Distributions of orbital elements. Squares (filled and open) represent
the currently observed orbital elements, with the open squares (with $a < $0.2 
AU) being candidates for significant tidal evolution. Triangles show the initial
orbital elements ($e_{initial}$ and $a_{initial}$) determined by integrating the
equations of tidal evolution backward in time to the formation of the planet, using $Q$ values that gave the best fit of the $e$ distribution to that of the other planets.}
\label{fig1}
\end{center}
\end{figure}

As the number of known extra-solar planetary orbits has increased,
interesting patterns are emerging. Figure \ref{fig1} shows the
semi-major axes $a$ and eccentricities $e$. Eccentricities of
extra-solar planets with $a >$ 0.2 AU average 0.3 and are broadly
distributed up to near 1. The distribution of eccentricities is fairly
uniform over $a$. For example, a Kolmogorov-Smirnov (K-S) test shows
that the $e$ distribution for $a$ between 0.2 and 1.0 AU matches that
for $a$ between 1.0 and 5.0 AU at the 96\% confidence level. However,
for close-in extra-solar planets (by which we mean $a <$ 0.2 AU), $e$
values are smaller, with an average of 0.09, but still large compared
to our solar system. For $a <$ 0.2 AU, the K-S test shows agreement at
only the 0.1\% level compared with planets further out.

\indent Here we consider the conventional idea that close-in planets
began with a distribution of $e$ similar to that of planets farther
out. Because the magnitude of tidal effects falls off very rapidly
with increasing $a$, \cite{Rasio96} suggested that tides could have
reduced $e$ for close-in planets and not for those farther out.

The implications of this hypothesis for planetary masses $M_p$, radii
$R_p$, and heating rates has been considered in previous studies
(\cite{Rasio96}; \cite{Trilling00}; \cite{Bodenheimer03}). Several
assumptions have been common. First, the tidal dissipation parameter
$Q_p$ for the planet is usually based on model-dependent estimates
based on the tidal evolution of the Jovian satellite system
(\cite{Yoder81}) or on physical theories of tidal dissipation
(\cite{Goldreich77}; \cite{Ogilvie04}). Second, effects of tides
raised on the star are only partially taken into account. Often the
orbit-circularizing effects of tides raised on the host star are
ignored, although they can be important. Third, some studies
(\textit{e.g.} \cite{Ford06}) assumed that orbital angular momentum is
conserved during tidal evolution, which is not true when tides on the
star are taken into account. Fourth, previous work ignored the strong
coupling of the tidal evolution of $e$ with that of $a$, generally
describing the circularization of an orbit in terms of an implicitly
exponential ``timescale''. In fact, the coupling with $a$ means that
evolution is more complex than a simple exponential damping. Moreover,
tidal evolution has resulted in significant inward migration of many
close-in planets.

In our study, we avoid these assumptions. We use the full tidal
evolution equations to determine possible past orbital change. 
The equations account for tides raised on the stars by the planet and on 
the planet by the star, each of which affects changes in both $e$ and $a$. We
employ conservative assumptions about $M_p$ and $R_p$, to test the
tidal circularization hypothesis.  And we consider a wide range of
possible $Q$ values, allowing us to determine which pair of $Q_p$ and
$Q_{*}$ values yields the most plausible evolution history.

These calculations also yield the corresponding past tidal heating
history for each planet. During the course of the tidal evolution,
tidal distortion of the figure of the planet can result in substantial
amounts of internal heating at the expense of orbital energy, so the
heating rate as a function of time is coupled to the evolution of the
orbit. In a typical case, tidal heating might have begun modest, but
then increased as tides reduced $a$. As the tides became stronger,
they would circularize the orbit, which in turn would shut down the
tidal heating mechanism. The relative strength and timing of these two
effects would determine a planet's history, typically with a gradual
increase in the heating rate followed by a decrease.

The thermal history of a planet is critical to determining its
physical properties. For example, studies of extra-solar planets have
considered the effects of heating on their radii, which can be
measured directly by transit observations. Heat sources that have been
considered in these models include the energy of planetary accretion
and radiation from the star, as well as tidal heating
(\cite{Bodenheimer03}; \cite{Mardling07}). In many cases the
theoretical predictions match the observations reasonably well
(\cite{Burrows07}). However, there are notable exceptions. HD 209458 b
has been observed by \cite{Knutson07} to have a radius of 1.32 Jupiter
radii ($R_J$), which is 10-20\% larger than predicted by theoretical
modeling \cite{Guillot05}. Similarly, HAT-P-1 b is 10-20\% larger than
predicted by theory (\cite{Bakos07a}). On the other hand, HAT-P-2 b seems to be
smaller than theory would predict (\cite{Bakos07b}), while for GJ 436 b,
theory predicts a radius consistent with observation (\cite{Gillon07}).

Theoretical models to date have not taken into account the history of
tidal heating for close-in planets, and of course those are the
planets most likely to have radii measurable by transits. The tidal
heating histories reported here and in \cite{Jackson08b}
provides motivation and a basis for construction of improved physical
models.

\section{Method}
To test the hypothesis that tides have been responsible for reducing
$e$, we numerically integrated the canonical tidal evolution equations
of \cite{Goldreich66} and \cite{Kaula68} backwards in time for all
close-in planets for which we have adequate information (see
\cite{Jackson08a} for details). For each planet, we began the
integration with the current best estimates of $e$ and $a$
($e_{current}$ and $a_{current}$) and integrated backwards over the
estimated age of the host star to find the orbital elements,
$e_{initial}$ and $a_{initial}$. We assumed tidal evolution dominated
the orbital evolution after the protoplanetary disk dissipated and
collisional effects became negligible, which probably happened only a
few Myr after the star's formation. Hence $e_{initial}$ and
$a_{initial}$ may describe the orbits at that time. We repeated the
integration for 289 combinations of $Q_p$ and $Q_{*}$, each $Q$
ranging from $10^4$ to $10^8$.

Our study involves a number of assumptions and approximations which
are detailed and discussed by \cite{Jackson08a}. In particular, our model 
assumes host stars rotate much more slowly than their close-in companion planet 
revolves, an assumption largely corroborated by observation (\cite{Trilling00}; 
\cite{Barnes2001}). As a result, tides raised on the star by the planet 
tend to decrease both $e$ and $a$, as do tides raised on the planet.
Our results will
inevitably need to be revisited and refined as improved data and
physical models become available. Of necessity, we considered only
planets for which the reported $e$ is non-zero, and for which there is
some estimate available for the age of the system. Even with these 
restrictions,
we can still study about 40\% of all known close-in planets. Stellar masses and
radii come from a variety of sources: \cite{Bakos07a}; \cite{Bakos07b}; 
\cite{DaSilva06}; \cite{Fischer05}; \cite{Moutou06}; \cite{Saffe06};
\cite{Gorda96}; and \cite{Takeda07}.  For planetary masses, we use the
radial-velocity minimum mass. For planetary radii, if $M_p >$ 0.3
Jupiter's mass, we fix $R_p$ = 1.2 $R_J$ since, for Jovian planets in
this range of mass, the radius is insensitive to mass
(\cite{Hubbard84}). This value is near the average radius of almost
all observed transiting extra-solar planets. For planets with $M_p <
$0.3 Jupiter's mass, we assume the planet has Jupiter's density and
scale $R_p$ accordingly. This assumption agrees fairly well with the
observed radius for GJ 436 b (\cite{Gillon07}; \cite{Deming07}). Where
$R_p$ and $M_p$ have been determined directly from transit
observations, we instead use those values.  Planetary data were taken
from a variety of sources: \cite{Bakos07a}; \cite{Bakos07b}; \cite{Butler06};
\cite{DaSilva06}; \cite{Deming07}; \cite{Gillon07}; \cite{Johnson06};
\cite{Knutson07}; \cite{Laughlin05}; \cite{Lovis06};
\cite{McArthur04}; \cite{Maness07}; \cite{Mayor04}; \cite{Moutou06};
\cite{Rivera05}; \cite{Udry02}; \cite{Valenti05}; \cite{Vogt05};
\cite{Wright06}; and \cite{Zucker04}.

Based upon the above model, we can also calculate the tidal heating
rate for planets undergoing tidal circularization. The tidal heating
of the planet results from the reduction of orbital energy (and hence
$a$) due to energy dissipation in the planet. By tracking changes in
$a$ due to tides raised on the planet throughout the process of tidal
circularization, we can estimate the tidal heating rate over the whole
lifetime of the planet (\cite{PCR79}). The magnitude and, indeed, the
very shape of the past heating curve over time depends sensitively
upon the assumed current orbital elements. Accordingly, we have
calculated multiple plausible heating curves for each planet,
corresponding to the range of observationally allowed values for
$e_{current}$ and $a_{current}$.

\begin{figure}[h]
\begin{center}
 \includegraphics[width=13cm]{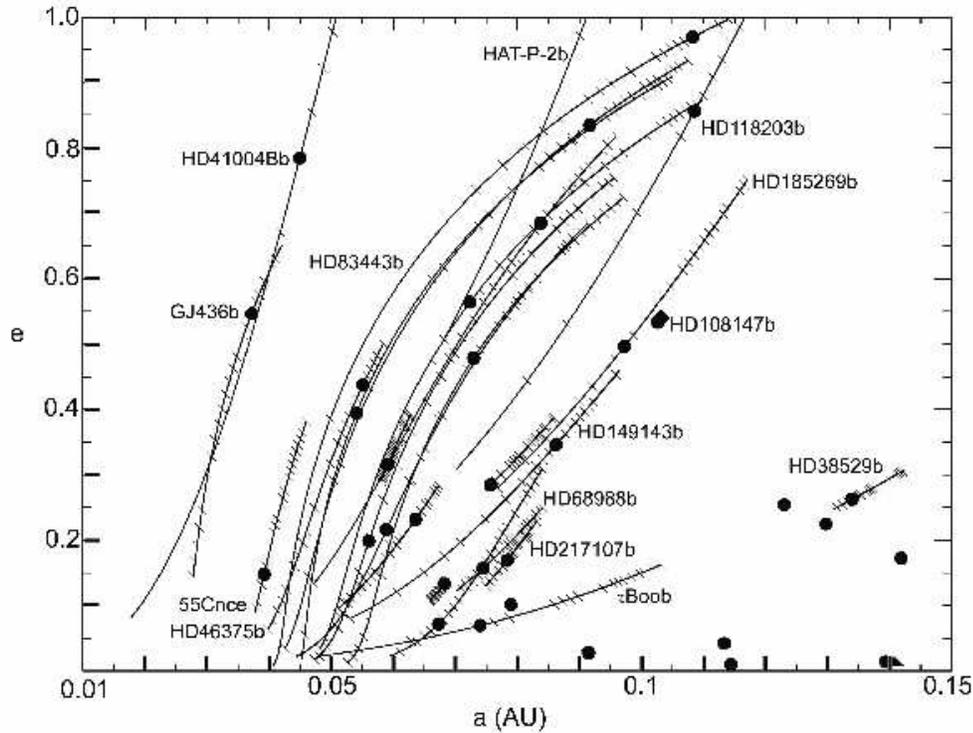}
 \caption{Tidal evolution of $e$ and $a$ for the sample of known close-in 
extra-solar planets using our best-fit values of $Q_* = 10^{5.5}$ and $Q_p = 
10^{6.5}$. Solid curves represent the trajectories of orbital evolution from 
current orbits (lower left end of each curve) backward in time (toward the upper
right). On the trajectories, tick marks are spaced every 500 Myr to indicate the
rate of tidal evolution. Tidal integrations were performed for 15 Gyr for all 
planets, but the filled circles indicate the initial values of orbital elements 
at the beginning of each planet's life. Due to space restrictions, most planets 
are not labeled, however they can be identified by the ($e$, $a$) values at the 
lower left end of each trajectory. \cite{Jackson08a} includes 
a table of 
$e_{current}$ and $a_{current}$ (Table 1) which can be used to identify planets 
in this figure.}
\label{fig2}
\end{center}
\end{figure}

\section{Orbital Evolution}
The 289 combinations of $Q_p$ and $Q_{*}$ that we tested gave a wide
variety of distributions of initial eccentricities for the close-in
planets. (Remember ``initial'' refers to the time, shortly after
formation, that a planet's orbital evolution begins to be dominated by
tides.) In addition to the current orbits (squares), Figure \ref{fig1}
shows the computed distributions of $e_{initial}$ and $a_{initial}$
(filled triangles) for the case of $Q_p$ = $10^{6.5}$ and $Q_{*}$ =
$10^{5.5}$. We compared the $e_{initial}$ distributions for the
close-in planets with the standard $e$ distribution observed for $a >
$ 0.2 AU. In this case, the agreement is excellent, with a K-S score
of 90\%.  These $Q$ values are well within plausible
ranges. Reasonable fits can also be obtained with other values of
$Q_{*}$ as long as $Q_p \sim10^{6.5}$. Other good fits (K-S $\sim$
70\%) have $Q_{*}$ = $10^{4.25}$ or $> 10^{6.75}$.

Figure \ref{fig2} shows the evolutionary track of $a$ and $e$ over
time for each of the planets, in the case of the best-fit $Q$
values. The current orbital elements are at the lower left end of each
trajectory in ($a$, $e$) space. These points correspond to empty
squares in Figure \ref{fig1}. The tick marks show the orbital elements
at intervals of 500 Myr, going back in time, for 15 Gyr, from the
present toward the upper right. Black dots have been placed at a point
representing the best age estimate for the planetary system. These
same points appear as the triangles in Fig. \ref{fig1}.

The evolutionary histories derived here include substantial changes in
semi-major axis coupled with the changes in eccentricity. For many
close-in planets, Figures \ref{fig1} and \ref{fig2} show that initial
$a$ values were significantly higher than the currently observed
values. These initial values of $a$ likely represent their locations
at the termination of gas disk migration in each early planetary
system. Given the extent of tidal migration for observed close-in planets, 
planets that formed inward of 0.04 AU could have subsequently fallen 
into their host stars. Consideration of such hypothesized scenarios will be the
subject of future work.

\begin{figure}[b]
\begin{center}
\includegraphics[width=13cm]{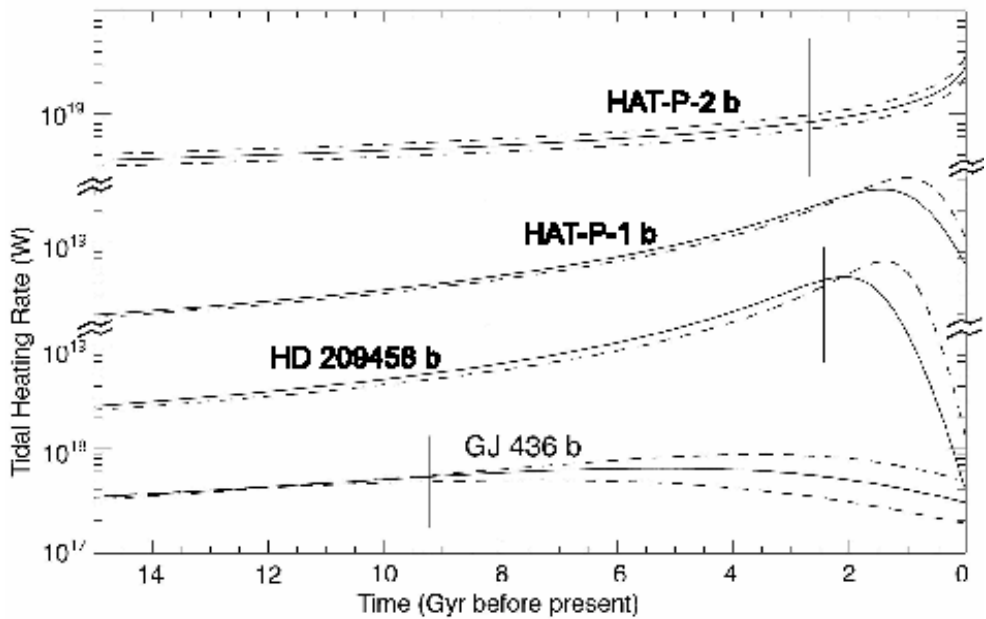}
 \caption{The tidal heating rates for planets HAT-P-2 b, HAT-P-1 b, 
HD 209458 b, and GJ 436 b as a function of time. The present time ($t = 0$) 
is at 
the right. The solid vertical line through each of the heating curves represents
the estimated time of formation for that planet. Note that the vertical scales 
has been shifted for HAT-P-2 b and HAT-P-1 b to make its curves more visible. 
(The vertical 
scale that corresponds to each curve is the scale intersected by that curve.) 
The solid curve for each planet is based on the current nominal eccentricity 
value. The dashed lines represent upper and lower limits on the heating, 
\textit{i.e.}\ they assume the extreme values of current $a$ and $e$ 
consistent with observations that give the maximum and minimum current tidal 
heating rates, respectively.  For HAT-P-1 b and HD 209458 b, observations could 
not exclude a current eccentricity of zero, so the lower bound on heating rates 
is formally zero. Hence in those cases only one dashed line is shown, 
representing the upper limit.} 
\label{fig3}
\end{center}
\end{figure}

\section{Tidal Heating}
Using the results of our tidal evolution calculations, we modeled the
past tidal heating rates for numerous planets. In Figure \ref{fig3},
we illustrate plausible heating histories for three interesting
examples: HD 209458 b, HAT-P-1 b and GJ 436 b, using $Q_p$ =
$10^{6.5}$ and $Q_{*}$ = $10^{5.5}$. Results for other planets are
presented in \cite{Jackson08b}. Note that in these examples the
planetary radii have been held constant, even though the changing
heating rates would most likely produce a changing radius.

For HD 209458 b, \cite{Burrows07} suggest that a heating rate of about
4 x $10^{19}$ W would be required to yield the observed planetary
radius, which is much larger than any allowable current tidal heating
rate.  However, the history plotted in Figure \ref{fig3} shows that
the required heating rate was available as recently as 1 billion years
ago. If the lag in the response of the planet to the heating rate were
on the order of a billion years, this heating rate may explain the
observed large radius. Such a lag seems reasonable based on the long
duration of the influence of heat of formation on the planet's radius
in the modeling by \cite{Burrows07}.

Like HD 209458 b, HAT-P-1 b's observed radius of 1.36 $R_J$
(\cite{Bakos07a}) is larger than expected from theoretical modeling
that did not include tidal heating (\cite{Guillot05}). Similar to HD
209458 b, Figure \ref{fig3} shows that its heating rate $\sim$ 1 Gyr
ago was substantially higher than the present tidal heating. For both
HD 209458 b and HAT-P-1 b, the substantial heating rate $\sim$ 3-4 x
$10^{19}$ W about 1 Gyr ago may help account for the discrepancy
between the large observed planetary radii and the predictions of
physical modeling.

In the case of HAT-P-2 b there also has been a substantial amount 
of tidal heating. The current heating rate is similar to the maximum rate 
attained by HD 209458 b and HAT-P-1 b, so again
we might expect a larger radius than predicted by theory that ignored tidal 
heating. In this case,
however, the measured radius is actually about 12\% smaller than predicted 
(\cite{Bakos07b}). Thus there
is a discrepancy between theory and observation even if tidal heating is 
neglected. The fact that there is likely a high rate of tidal dissipation makes 
the problem even worse. On the other hand, a key factor in the reconciliation 
may be that, while the current tidal
heating rate is high and increasing, in the recent past the heating rate was 
much lower. HAT-P-2 b is still on the increasing part of the heating curve, 
which is unusual among planets considered
here, most of which have passed their peaks. The fact that the heating rate was 
several times smaller a billion years ago than it is now may help explain the 
small radius.\\

Figure \ref{fig3} also illustrates tidal heating histories for GJ 436
b.  This planet has a measured radius consistent with theoretical
models, independent of tidal heating (\cite{Gillon07}). The tidal
heating history shown in Figure \ref{fig3} is consistent with that
result. Compared with the previous two cases, the maximum heating rate
was two orders of magnitude less, small enough perhaps not to affect
the radius.

\section{Discussion}
This investigation supports the hypothesis that tidal interactions
between a star and a planet are responsible for the relatively small
$e$ values of close-in planets (\cite{Rasio96}), although our
calculations introduce important corrections to previous
studies. Because even close-in planets evidently formed with much
larger $e$, the processes that governed their early dynamics were
probably similar to other planets. A plausible mechanism for producing
the initial $e$ distribution of extra-solar planets is planet-planet
scattering (\cite{RF96}; \cite{WM96}), although some modification to the
original model are needed (\cite{Barnes07}).

If the tidal circularization hypothesis is borne out, then our study
also provides constraints on $Q$ values. We find that agreement
between eccentricity distributions requires $Q_p \sim10^{6.5}$, with
$Q_{*} \sim10^{4.25}$, $10^{5.5}$ (best fit), or $10^{6.75}$. This
$Q_p$ is in good agreement with other constraints (\cite{Yoder81};
\cite{Ogilvie04}), and this $Q_{*}$ agrees well with studies of binary
star circularization (\cite{Mathieu94}). Of course, it is likely that
individual planets and stars have unique $Q$ values, owing to
variation in their internal structures. These $Q$ values are only
meant to be representative values, good for the population as a
whole. Corrections to our tidal model might also result in different
suitable $Q$ values.

Significant reductions in semi-major axes have accompanied the changes
in eccentricity, with important implications. First, models of
protoplanetary migration in the primordial gas disk need not carry
``hot Jupiters'' in as far as their current positions. \cite{Lin96}
proposed that migration in the gas disk halted near the inner edge of
the disk, a boundary determined by clearing due to the host star's
magnetosphere. Our results show that the inner edge was probably
farther out than indicated by the current semi-major axes of the
planets, which were only reached during tidal migration long after the
nebula had dissipated. In order to evaluate where migration due to the
gas disk halted, (and thus where the inner edge of the nebula was)
models should account for the subsequent tidal evolution.

The tidal changes in orbital semi-major axes also have implications
for observations of planetary transits, such as surveys of young open
galactic clusters (\cite{Bramich05}; \cite{vonBraun05};
\cite{Burke06}). The probability to observe a planetary transit increases 
for smaller semi-major axes, but decreases as orbits become more circular 
(\cite{Borucki84}; \cite{Barnes2007}). Tidal evolution means that the 
probability of an observable transit depends on a star's age, but the exact 
relation depends on the particular evolutionary path through ($a$, $e$) space. 
As our understanding of the statistics of tidal evolution paths improves, the 
observed frequency of transits in the field and in open clusters may eventually 
help to constrain planetary formation scenarios, distinguishing, for example, 
between the relative roles
of embedded migration and of gravitational scattering, which set up the initial 
conditions for tidal evolution. Transit statistics may not yet be refined 
enough to be sensitive to detect this effect (\cite{Pepper06}), but such 
systematic effects may show up in future surveys.

The tidal heating calculations here suggest that past tidal heating
may well have played an important role in the evolution of the
physical properties of many extra-solar planets, specifically the
planetary radius. We caution that the specific calculations displayed
here depend on numerous assumptions and several uncertain
parameters. The heating rates correspond to the orbital evolution
trajectories computed by \cite{Jackson08a}, and various caveats are
discussed in detail there. It is quite likely that the actual thermal
history of any particular planet was different to some degree from
what we show here. In particular, when analyzing radial-velocity
observations to solve for a planet's orbit, it is difficult to rule
out a completely circular orbit, in which case the tidal heating rate
would have been zero. However, the unavoidable point is that past
tidal heating may be significant for many planets and should be
considered as a factor in theoretical modeling of physical properties
of exoplanets.

For every planet whose tidal evolution we modeled, we have calculated
corresponding tidal heating histories (\cite{Jackson08b}). For the
cases presented here, see Fig.\ \ref{fig3}, we see that past tidal
heating may provide a previously unconsidered source of heat for
planets with larger-than-predicted radii. However, it may make things
worse in cases where measured radii seemed to fit the current
models. Theoretical models of tidal evolution and planetary interiors
will generally need to be adjusted and improved so as to yield a match
between predicted and observed planetary radii.

To conclude, we find that the distribution of orbital eccentricities for 
exoplanets was once strikingly uniform across all semi-major axes. By varying
tidal dissipation parameters, we can match the original
distribution of close-in planetary eccentricities to that of planets far from
their host star for stellar and planetary $Q$'s $\sim10^{5.5}$ and 
$\sim10^{6.5}$, which are consistent with previous estimates. After the 
formation of the close-in exoplanets, tides raised on the host star and on the 
planet acted over Gyrs to reduce orbital eccentricities and semi-major axes of 
the close-in exoplanets. This reduction in $e$ as well as $a$
has important implications for the thermal histories of close-in exoplanets and 
for transit studies.

We also find that tidal heating in the past was significantly larger
than current heating. For example, about 1 Gyr ago, HD 209458 b may have 
undergone tidal heating 100 times the present value. This substantial heating
may help resolve the mystery of the anomalously large radii observed for many
transiting planets today. If the lag in response of the planetary radii to
tidal heating is of order a Gyr, then past tidal heating must be included in
models of exoplanetary radii. Previous studies suggest such a lag is reasonable.
However, further studies are required to elucidate this effect.

\bibliographystyle{plain}

\begin{thebibliography}{}
\bibitem[Bakos \etal\ (2007a)]{Bakos07a}
{Bakos, P., \textit{et al}.} 2007a,
\textit{ApJ}, 656, 552.

\bibitem[Bakos \etal\ (2007b)]{Bakos07b}
{Bakos, P., \textit{et al}.} 2007b,
\textit{ApJ}, 670, 826.

\bibitem[Barnes (2001)]{Barnes2001}
{Barnes} 2001,
\textit{ApJ}, 561, 1095.

\bibitem[Barnes \& Greenberg (2007)]{Barnes07}
{Barnes, R., \& Greenberg, R.} 2007,
\textit{ApJL}, 659, L53.

\bibitem[Barnes (2007)] {Barnes2007}
{Barnes, J.} 2007,
\textit{PASP}, 119, 986.

\bibitem[Bodenheimer \etal\ (2003)]{Bodenheimer03}
{Bodenheimer, P., Laughlin, G., \& Lin, D.N.C} 2003,
\textit{ApJ}, 592, 555.

\bibitem[Borucki \& Summers (1984)]{Borucki84}
{Borucki, W.J., \& Summers, A.L.} 1984,
\textit{Icarus}, 58, 121.

\bibitem[Bramich \etal\ (2005)]{Bramich05}
{Bramich, D. M. \textit{et al}.} 2005,
\textit{Mon. Not. R. Astron. Soc.}, 359, 1096.

\bibitem[Butler \etal\ (2006)]{Butler06}
{Butler, R.P. \textit{et al}.} 2006,
\textit{ApJ}, 646, 505.

\bibitem[Burke \etal\ (2006)]{Burke06}
{Burke, C.J. \textit{et al}.} 2006,
\textit{ApJ}, 132, 210.

\bibitem[Burrows \etal\ (2007)]{Burrows07}
{Burrows, A., Hubeny, I., Budaj, J., \& Hubbard, W.B.} 2007,
\textit{ApJ}, 661, 514.

\bibitem[Da Silva \etal\ (2006)]{DaSilva06}
{Da Silva, R. \textit{et al}.} 2006,
\textit{A\&A}, 446, 717.

\bibitem[Deming \etal\ (2007)]{Deming07}
{Deming, D. \textit{et al}.} 2007,
\textit{ApJL}, 667, L199.

\bibitem[Fischer \& Valenti (2005)]{Fischer05}
{Fischer, D. \& Valenti, J.} 2005,
\textit{ApJ}, 622, 1102.

\bibitem[Ford \& Rasio (2006)]{Ford06}
{Ford, E. \& Rasio, F.} 2006,
\textit{ApJ}, 638, L45.

\bibitem[Gillon \etal\ (2007)]{Gillon07}
{Gillon, M. \textit{et al}.} 2007,
\textit{A\&A}, 472, L13.

\bibitem[Goldreich \& Nicholson (1977)]{Goldreich77}
{Goldreich, P. \& Nicholson, P.} 1981,
\textit{Icarus}, 30, 301.

\bibitem[Goldreich \& Soter (1966)]{Goldreich66}
{Goldreich, P. \& Soter, S.} 1966,
\textit{Icarus}, 5, 375.

\bibitem[Gorda \& Svechnikov (1996)]{Gorda96}
{Gorda, S. \& Svechnikov, M. A.} 1996,
\textit{Astron. Reports}, 42, 793.

\bibitem[Guillot (2005)]{Guillot05}
{Guillot, T.} 2005,
\textit{Ann. Rev. Earth and Planet. Sci.}, 33, 493.

\bibitem[Hubbard (1984)]{Hubbard84}
{Hubbard, W.B.} 1984,
Planetary Interiors, (New York: Van Nostrand Reinhold Co).

\bibitem[Jackson \etal\ (2008a)]{Jackson08a}
{Jackson, B., Greenberg, R., \& Barnes, R.} 2008a
\textit{ApJ}, accepted

\bibitem[Jackson \etal\ (2008b)]{Jackson08b}
{Jackson, B., Greenberg, R., \& Barnes, R.} 2008b, \textit{ApJ}, submitted.

\bibitem[Johnson \etal\ (2006)]{Johnson06}
{Johnson, J.A. \textit{et al}.} 2006
\textit{ApJ}, 652, 1724.

\bibitem[Kaula (1968)]{Kaula68}
{Kaula, W.} 1968,
An Introduction to Planetary Physics, Wiley, NY.

\bibitem[Knutson \etal\ (2007)]{Knutson07}
{Knutson, H., Charbonneau, D., Noyes, R., Brown, T., \& Gilliland, R.} 2007,
\textit{ApJ}, 655, 564.

\bibitem[Laughlin \etal\ (2005)]{Laughlin05}
{Laughlin, G. \textit{et al}.} 2005
\textit{ApJL}, 629, L121.

\bibitem[Lin \etal\ (1996)]{Lin96}
{Lin, D.N.C. \textit{et al}.} 1996,
\textit{Nature}, 380, 606.

\bibitem[Lovis \etal\ (2006)]{Lovis06}
{Lovis, C. \textit{et al}.} 2006
\textit{Nature}, 441, 305.

\bibitem[Maness (2007)]{Maness07}
{Maness, H.L. \textit{et al}.} 2007,
\textit{PASP}, 119, 90.

\bibitem[Mardling (2007)]{Mardling07}
{Mardling, R.} 2007,
\textit{Mon. Not. R. Astron. Soc.}, in press.

\bibitem[Weidenschilling \& Marzari (1996)]{WM96}
{Weidenschilling, S.J. \& Marzari, F.} 1996,
\textit{Nature}, 384, 619. 

\bibitem[Mathieu (1994)]{Mathieu94}
{Mathieu, R.} 1994,
\textit{Annu. Rev. Astron. Astrophys.}, 32, 465.

\bibitem[Mayor (2004)]{Mayor04}
{Mayor, M. \textit{et al}.} 2004,
\textit{A\&A}, 415, 391.

\bibitem[McArthur \etal\ (2004)]{McArthur04}
{McArthur, B. \textit{et al}.} 2004,
\textit{ApJL}, 614, L81.

\bibitem[Moutou \etal\ (2006)]{Moutou06}
{Moutou, C. \textit{et al}.} 2006,
\textit{A\&A}, 458, 327.

\bibitem[Ogilvie \& Lin (2004)]{Ogilvie04}
{Ogilvie, G. \& Lin, D.N.C.} 2004,
\textit{ApJ}, 610, 477.

\bibitem[Peale \etal\ (1979)]{PCR79}
{Peale, S.J., Cassen, P. \& Reynolds, R.T.} 1979,
\textit{Science}, 203, 892.

\bibitem[Pepper \& Gaudi (2006)]{Pepper06}
{Pepper, J. \& Gaudi, B.S.} 2006,
\textit{Acta Astronomica}, 56, 183.

\bibitem[Rasio \& Ford (1996)]{RF96}
{Rasio, F.A. \& Ford, E.B.} 1996,
\textit{Science}, 279, 954.

\bibitem[Rasio \etal\ (1996)]{Rasio96}
{Rasio, F. A., Tout, C.A., Lubow, S.H., \& Livio, M.} 1996,
\textit{ApJ}, 470, 1187.

\bibitem[Rivera \etal\ (2005)]{Rivera05}
{Rivera, E. \textit{et al}.} 2005,
\textit{ApJ}, 634, 625.

\bibitem[Saffe \etal\ (2006)]{Saffe06}
{Saff\`{e}, C. \textit{et al}.} 2006,
\textit{A\&A}, 443, 609.

\bibitem[Takeda \etal\ (2007)]{Takeda07}
{Takeda, G. \textit{et al}.} 2007,
\textit{ApJS}, 168, 297.

\bibitem[Trilling (2000)]{Trilling00}
{Trilling, D.} 2000,
\textit{ApJL}, 537, L61.

\bibitem[Udry \etal\ (2002)]{Udry02}
{Udry, S. \textit{et al}.} 2002,
\textit{ApJ}, 634, 625.

\bibitem[Valenti \& Fischer (2005)]{Valenti05}
{Valenti, J. \& Fischer, D.} 2005,
\textit{ApJ}, 159, 141.

\bibitem[Vogt \etal\ (2005)]{Vogt05}
{Vogt, S. \textit{et al}.} 2005,
\textit{ApJ}, 632, 638.

\bibitem[von Braun \etal\ (2005)]{vonBraun05}
{von Braun, K. \textit{et al}.} 2005,
\textit{PASP}, 117, 141.

\bibitem[Wright \etal\ (2006)]{Wright06}
{Wright, J.T. \textit{et al}.} 2007,
\textit{ApJ}, 657, 533.

\bibitem[Yoder \& Peale (1981)]{Yoder81}
{Yoder, C. \& Peale, S.} 1981,
\textit{Icarus}, 47, 1.

\bibitem[Zucker \etal\ (2004)]{Zucker04}
{Zucker, S. \textit{et al}.} 2004,
\textit{A\&A}, 426, 695.

\end{thebibliography}

\end{document}